\documentclass[12pt]{article}
\usepackage[dvips]{color}
\usepackage{epsfig}
\usepackage{amsmath}
\usepackage{graphicx}
\begin{document}
\title{\bf The orbit method solution for the deformed three coupled scalar fields}

\author{J.  Sadeghi$^{a}$\thanks{Email:pouriya@ipm.ir}\hspace{1mm} , A. R. Amani $^b$
\thanks{Email:a.r.amani@iauamol.ac.ir}\hspace{1mm} and  A. Pourdarvish$^{c}$ \thanks{Email:a.pourdarvish@umz.ac.ir}
  \\ $^a$
{\small {\em  Sciences Faculty, Department of Physics, Mazandaran University,}}\\
        {\small {\em P.O.Box 47415-416, Babolsar, Iran}}\\
$^b${\small {\em Department of Physics, Islamic Azad University-Ayatollah Amoli Branch,}}\\
{\small {\em P. O. Box 678, Amol, Mazandaran, Iran }} \\
$^c${\small {\em  Sciences Faculty, Department of Statistics, Mazandaran University}}\\
        {\small {\em P.O.Box 47415-416, Babolsar, Iran}}}

\maketitle

\begin{abstract}
\noindent  In this work, we present a deformed solutions starting
from systems   of three coupled scalar fields with super-potential
$W(\phi_1, \phi_2, \phi_3)$ by orbit method. First, we deform the
corresponding super-potential and obtain defect solutions.  It is
shown that how to construct new models altogether with its defect
solutions in terms of the non-deformed model. Therefore , we draw
the graph of super-potential and different fields in terms of $x.$
So we observe that the graphs for deformed and non -
deformed cases are changed by the scale.\\

 \end{abstract}
\section{Introduction}

As we know the defect structures exist in different branches of
physics, such as domain wall, kinks, vertices, monopoles, condensed
matter and string theory.  In higher  space-times  dimensions the
defect structures are generated  by real scalar fields such that the
single real scalar field produce just single defect as kink-like and
the double Sin-Gordon model may create  two different defects. On
the other hand, models containing two or more real scalar fields can
give rise to at least two other classes of systems that produce
defect that engender internal structure and those that support
junctions of defects. Also two and three scalar fields describe the
regular hexagonal network, Higgs model [1-5] and bent brane in five
dimensions [6].  In other hand for field theories involving two and
three real
 scalar fields, the mathematical problem concerning the integrability of equations of motion is  much harder,
 as one deals with a system of two coupled second order nonlinear ordinary differential equations.
 Also the configuration space shows a distribution of minima that
 allows for a number of topological sectors.  One way of simplifying the problem is to consider potentials
 belonging to large class corresponding to bosonic sector of
 supersymmetric theories. Such systems can be studied by
 super-potential. This super-potential lead us to consider all
 second order equations in terms of first order. All
 Bogomol'nyi-Prasad-Sommerfield (BPS)
 configuration can be described by this first order equation. For
 models with three interacting components, the solutions on each
 topological sector determine orbits in the configuration space,
 which can be expressed as a constraint equation $C(\chi_1, \chi_2,
 \chi_3)=0.$ The equations arising for three fields in deformation
 procedure  more complicated then two and single field, so
 three deformation functions are required. It is difficult to solve
 this deformation equation, then  we restrict the orbit in field
 space. Therefore, by deforming the first order equations for a three
 coupled field system we need to  impose  the orbit constraint.
As we know the deformation method  for  the  single and two coupled
scalar fields are discussed by Bazeia and et al.  But we  are going
to consider three coupled scalar fields in (1,1) dimensions in flat
space-time , where metric is $\eta_{\mu\nu}=(1,-1)$ [1-4]. The first
- order differential equation and the static solutions such as
topological solutions  for stable states has been discussed by
 Refs. [7,8].\\
In present paper, we  use deformation procedure to obtain defect
solutions. This deformation plays an  important role to investigate
the energy  of systems. For example in   cosmological model, the
energy density, pressure and  equation of states can be controlled by the deformation method on the fields .\\
The above pretext give us motivation to discuss the deformation
procedure. So the outline of the paper follows. In the next section
three coupled scalar fields with an example is discussed and the
analysis of the solution without deformation is shown. The
deformation procedure for the three coupled scalar field with
diagram is discussed in Sec.3. Also we compare two diagrams  and
show that the variations of the fields with respect to coordinates
are changed  by scale.

\section{Three coupled scalar fields system}
 Lagrangian systems described by coupled scalar fields are gaining renewed attention.
  In the case of real fields, in particular,
 the work presented in [4] has introduced a specific class of systems of two coupled real scalar
 fields.\\
  Let us to start with the lagrangian density with three coupled system,
\begin{equation}\label{T1}
L=\frac{1}{2}\partial_{\mu}\phi_1\partial^{\mu}\phi_1+
\frac{1}{2}\partial_{\mu}\phi_2\partial^{\mu}\phi_2+\frac{1}
{2}\partial_{\mu}\phi_3\partial^{\mu}\phi_3-U(\phi_1,\phi_2,\phi_3),
\end{equation}
 where $U=U(\phi_1,\phi_2,\phi_3)$ is in general a non-linear function of the three fields  $\phi_1$, $\phi_2$ and $
 \phi_3.$ Here we are using natural units, and so $\hbar=1$  and the metric is such that $x^{\mu}=(t, x)$
 and $x_{\mu}=(t, -x).$\\
  By using  Euler-Lagrange equations, one can  obtain
the equations of motion as follows;

\begin{eqnarray}\label{T2}
\frac{\partial^{2}\phi_1}{\partial x^{2}}=\frac{d U}{d \phi_1}\nonumber\\
\frac{\partial^{2}\phi_2}{\partial x^{2}}=\frac{d U}{d \phi_2}\nonumber\\
\frac{\partial^{2}\phi_3}{\partial x^{2}}=\frac{d U}{d \phi_3}.
\end{eqnarray}
In order to obtain  solution for the equation  (2), we define
super-potential function  $W=W(\phi_1,\phi_2,\phi_3)$ such that one
may write the potential in terms of super potential,

\begin{eqnarray}\label{T3}
U(\phi_1,\phi_2,\phi_3)=\frac{1}{2}W_{\phi_1}^2+\frac{1}{2}W_{\phi_2}^2+\frac{1}{2}W_{\phi_3}^2,
\end{eqnarray}
where $W=W(\phi_1,\phi_2,\phi_3)$ is a smooth function and we have,

\begin{eqnarray}\label{T2}
W_{\phi_i}=\frac{\partial W}{\partial {\phi_i}},\qquad i=1,2,3.
\end{eqnarray}
The energy spectrum associated with these configurations could be
written as [9-10],
\begin{eqnarray}
E=\frac{1}{2}\int_{-\infty}^{+\infty}dx\sum_{i=1}^3\left[(\frac{d\phi_i}{dx})^2+W_{\phi_i}^2\right],
\end{eqnarray}
we can rewrite  it also as  following,
\begin{eqnarray}
E=E_{BPS}+\frac{1}{2}\int_{-\infty}^{+\infty}dx\sum_{i=1}^3\left[\frac{d\phi_i}{dx}-W_{\phi_i}\right]^2.
\end{eqnarray}
Here also we set the BPS energy,
\begin{eqnarray}
E_{BPS}=|\Delta {W}|= |W({\phi_i}(\infty)) - W({\phi_i}(-\infty))|.
\end{eqnarray}
This procedure shows that the energy is minimized to,
\begin{eqnarray}
E=E_{BPS}.
\end{eqnarray}
As we have already learned from [4], we impose conditions,
\begin{eqnarray}\label{T3}
\frac{d \phi_i}{d x}=W_{\phi_i},\qquad i=1,2,3.
\end{eqnarray}
In this case we see that the energy gets to its lower bound
$E_{BPS}$, and the above first - order equation (9) solve the
corresponding equations of motion (2).

Now we consider special example of three coupled scalar fields which
describes the regular hexagonal network.  Let us consider the system
defined by the following super-potential,
\begin{equation}
W(\phi_1,\phi_2,\phi_3)=\phi_1-\frac{\phi_1^{3}}{3}-r\phi_1(\phi_2^{2}+\phi_3^{2}),
\end{equation}
In this case the first - order equations become,

\begin{eqnarray}
\frac{d\phi_1}{dx}&=&1-\phi_1^{2}-r(\phi_2^{2}+\phi_3^{2})\nonumber\\
\frac{d\phi_2}{dx}&=&-2r\phi_1\phi_2\nonumber\\
\frac{d\phi_3}{dx}&=&-2r\phi_1\phi_3\nonumber\\.
\end{eqnarray}
As we know the  exact solution for the  above equations is not
clear, so we  shall use the following  elliptical orbit procedure
[6,9,11],
\begin{eqnarray}
\phi_1^{2}+\frac{\phi_2^{2}}{\frac{1}{r}-2}+\frac{\phi_3^{2}}{\frac{1}{r}-2}=1,
\end{eqnarray}
finally  three - field static solution for the system (11) is,
\begin{eqnarray}
\phi_1(x)&=&\pm\tanh(2rx)\nonumber\\
\phi_2(x)&=&\pm\sqrt{\frac{1}{r}-2} \cos(\theta) sech(2rx)\nonumber\\
\phi_3(x)&=&\pm\sqrt{\frac{1}{r}-2} \sin(\theta)
sech(2rx)\nonumber\\,
\end{eqnarray}

So we have a three- field model and its general orbit equation depending on two parameters
$r$ and  $\theta$ (the arbitrary phase). \\
We recall that there are several orbits for solving the  equation
(11). But here we shall consider the condition $0<r<\frac{1}{2}$ and
the orbit
equation (12).\\

By putting equation (13) in  (10)  and (3) the corresponding
super-potential and potential  in terms of $x$ are given by,
\begin{eqnarray}
W(x)=\frac{2}{3}tanh(2rx)[1+(3r-1)sech^2(2rx)],
\end{eqnarray}

\begin{eqnarray}
U(x)=2rsech^2(2rx)[(3r-1)sech^2(2rx)+1-2r].
\end{eqnarray}

\section{Deformation procedure and orbit solution}
We are going to apply deformation procedure for three  couple scalar
fields. As we know   the  deformation method for two and single
field  were  discussed  in   Ref.s. [9, 12, 13].\\ First we
transform three initial scalar fields $\phi_1(x)$, $\phi_2(x)$ and
$\phi_3(x)$ into the form of scalar fields $\chi_1(x)$, $\chi_2(x)$
and $\chi_3(x)$ respectively.\\ In order to deform three scalar
fields we   introduce the following deformed function;
\begin{eqnarray}
\phi_i(x)=f_i(\chi_i),\qquad i=1,2,3,
\end{eqnarray}
where non - deformed function $\phi_i(x)$ and deformed function
$\chi_i(x)$ are differentiable and invertible. We can write the
inverse function as $\chi_i(x)=f^{-1}_i(\phi_i)$ for $i=1, 2, 3$.
For simplicity the  deformed functions are just function of single
field. The essential condition for the deformation functions are
given by [12],
\begin{eqnarray}
\left[\frac {d f_1(\chi_1)}{d \chi_1}=\pm\frac {d f_2(\chi_2)}{d
\chi_2}=\pm\frac {d f_3(\chi_3)}{d \chi_3}\right]_{orbit}.
\end{eqnarray}
which is just the orbit-based deformation procedure. Also, we note
that in Ref. [12] they applied this procedure in two fields system
and we will apply for three fields system where
$\phi_1(x)=f_1(\chi_1)$, $\phi_2(x)=f_2(\chi_2)$ and
$\phi_3(x)=f_3(\chi_3)$.

This equation shows  that the variation of  non-deformed functions
are equivalent  to the  deformed functions. Finally the deformed
super-potential $\mathcal{W}(\chi_{1},\chi_2,\chi_3)$ from eqs. (4)
and (9)  will be as;
\begin{eqnarray}
\mathcal{W} = \int{\frac {W_{\phi_i}}{\frac{d f_i(\chi_i)}{d
\chi_i}}}d\chi_i.
\end{eqnarray}
With the  help  of essential  condition and eq. (3) we have,
\begin{eqnarray}
\mathcal{U} = \frac {U}{({\frac{d f_i(\chi_i)}{d \chi_i})^2}},
\end{eqnarray}
where $\mathcal{U}$ deformed potential in terms of  deformed
super-potential is,
\begin{eqnarray}
\mathcal{U} = \sum_{i=1}^{3} \frac {1}{2}\mathcal{W}_{\chi_i}^2.
\end{eqnarray}
 The energy of deformed defects can be  written by following expression,
\begin{eqnarray}
\mathcal{E}_{BPS}= \Delta\mathcal{W}|_{-\infty}^{+\infty},
\end{eqnarray}
where $\mathcal{E}$ is deformed  BPS energy .\\\\

Now we apply  deformation method for super-potential (10). To start,
we  shall introduce the  deformed function $\chi_1$,

\begin{eqnarray}
\phi_1=f_1(\chi_1)=\tan(\chi_1),
\end{eqnarray}
and  from  eq. (9) $\chi_1$ is given by,
\begin{eqnarray}
\chi_1=arctan(\phi_1).
\end{eqnarray}
By using the following expression $$
\frac{\phi_{3}}{\phi_{2}}=c=\tan \theta$$ the  orbit equation (12)
will be as,
\begin{eqnarray}
\phi_1^2+\frac{(1+c^2)}{\frac{1}{r}-2}\phi_2^2=1.
\end{eqnarray}
In order to obtain the deformation fields $\chi_2$ and $\chi_3$, we
use  Eqs. (16) and (17), so we  shall have;
\begin{eqnarray}
\chi_2=\int\frac{d\chi_1}{d\phi_1}d\phi_2, \qquad
\chi_3=\int\frac{d\chi_1}{d\phi_1}d\phi_3.
\end{eqnarray}
Here we use equations (23), (24) and (13) the corresponding
deformed fields $\chi_2$ and $\chi_3$  are respectively;
\begin{eqnarray}
\chi_2=\sqrt{\frac{1}{2r}-1}  cos (\theta)
arctanh(\frac{1}{\sqrt{2}}sech(2rx)),
\end{eqnarray}

\begin{eqnarray}
\chi_3=\sqrt{\frac{1}{2r}-1}  \sin (\theta)
arctanh(\frac{1}{\sqrt{2}}sech(2rx)).
\end{eqnarray}

The  deformation functions  also will be the  following;

\begin{eqnarray}
\phi_2=f_2(\chi_2)=\sqrt{2(\frac{1}{r}-2)}  \cos (\theta)
tanh(\frac{sec(\theta)}{\sqrt{\frac{1}{2r}-1}}\chi_2),
\end{eqnarray}
\begin{eqnarray}
\phi_3=f_3(\chi_3)=\sqrt{2(\frac{1}{r}-2)}  sin (\theta)
tanh(\frac{\csc(\theta)}{\sqrt{\frac{1}{2r}-1}}\chi_3).
\end{eqnarray}
  The  deformed super-potential and  deformed potential from
equations (18) and (19) are given by;
\begin{eqnarray}
\mathcal{W}=rtanh(4rx),
\end{eqnarray}
and
\begin{eqnarray}
\mathcal{U}=\frac{2rsech^2(2rx)\left[(3r-1)sech^2(2rx)+1-2r\right]}{(1+tanh^2(2rx))^2}.
\end{eqnarray}
Finally we can say the  topological solutions of non-deformed and
deformed equations   are compared together by drawing their graphs
see  Figs. (1) and (2).
\begin{figure}[th]
\centerline{\epsfig{file=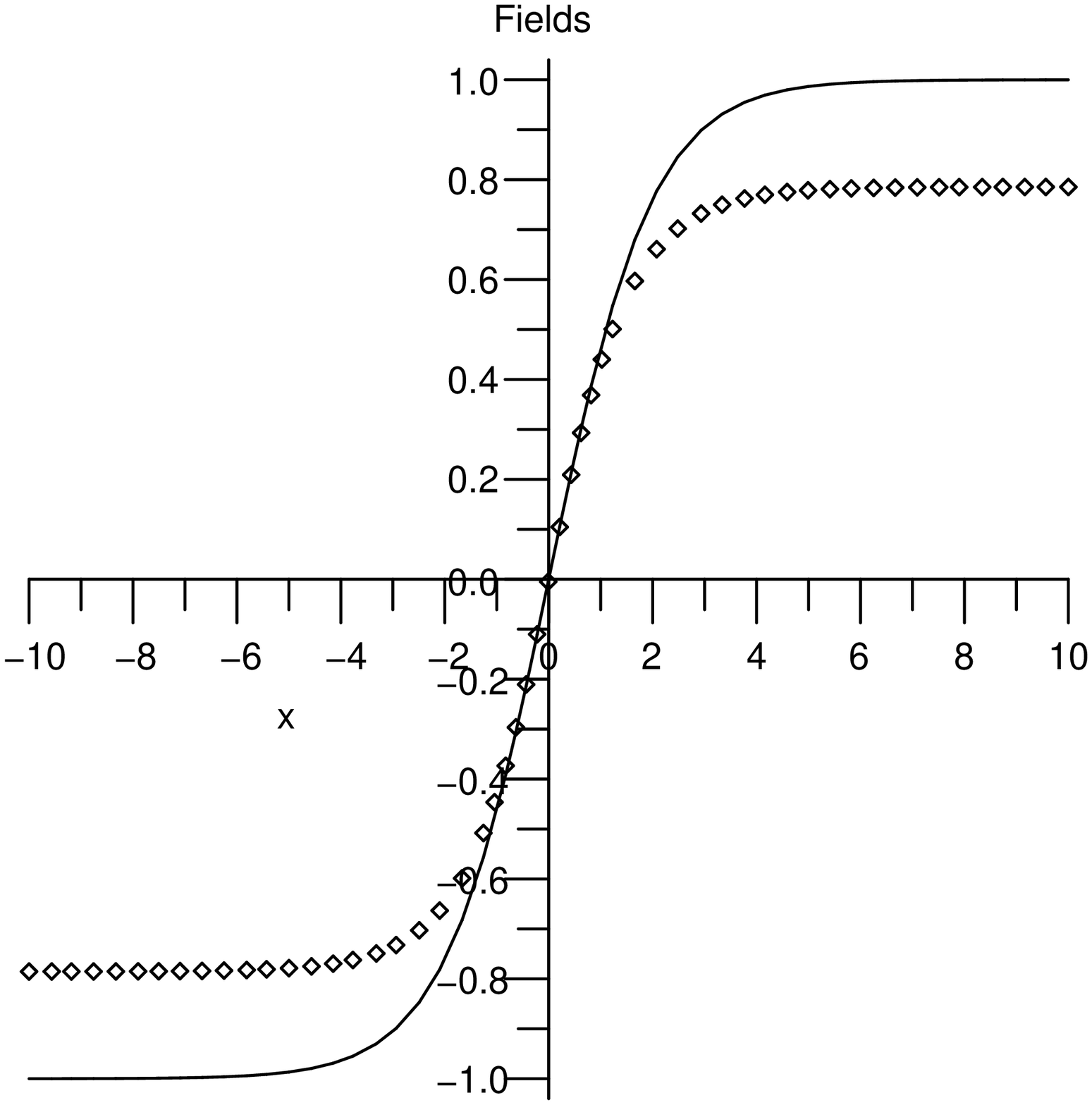,scale=.3}\epsfig{file=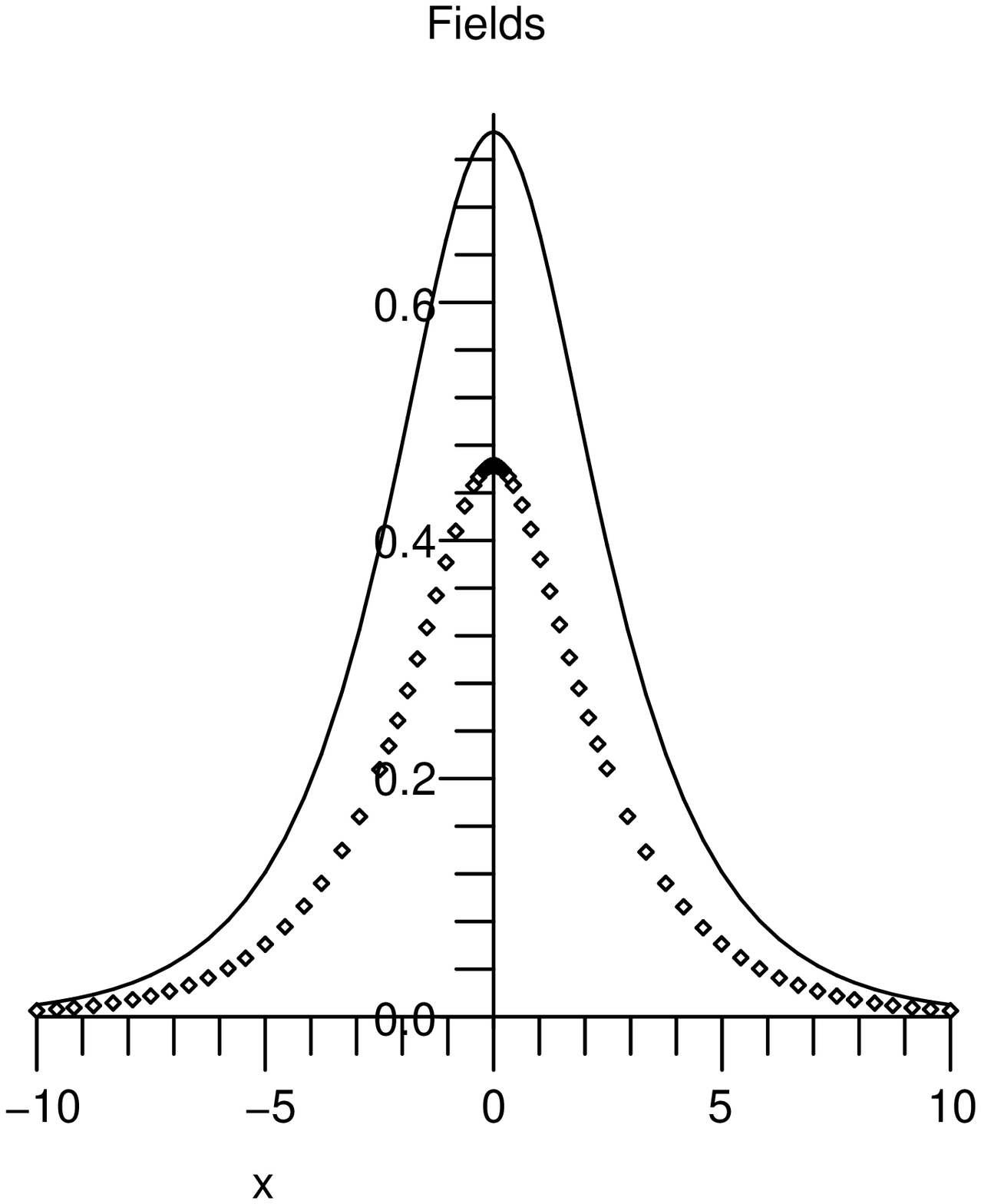,scale=.3}\epsfig{file=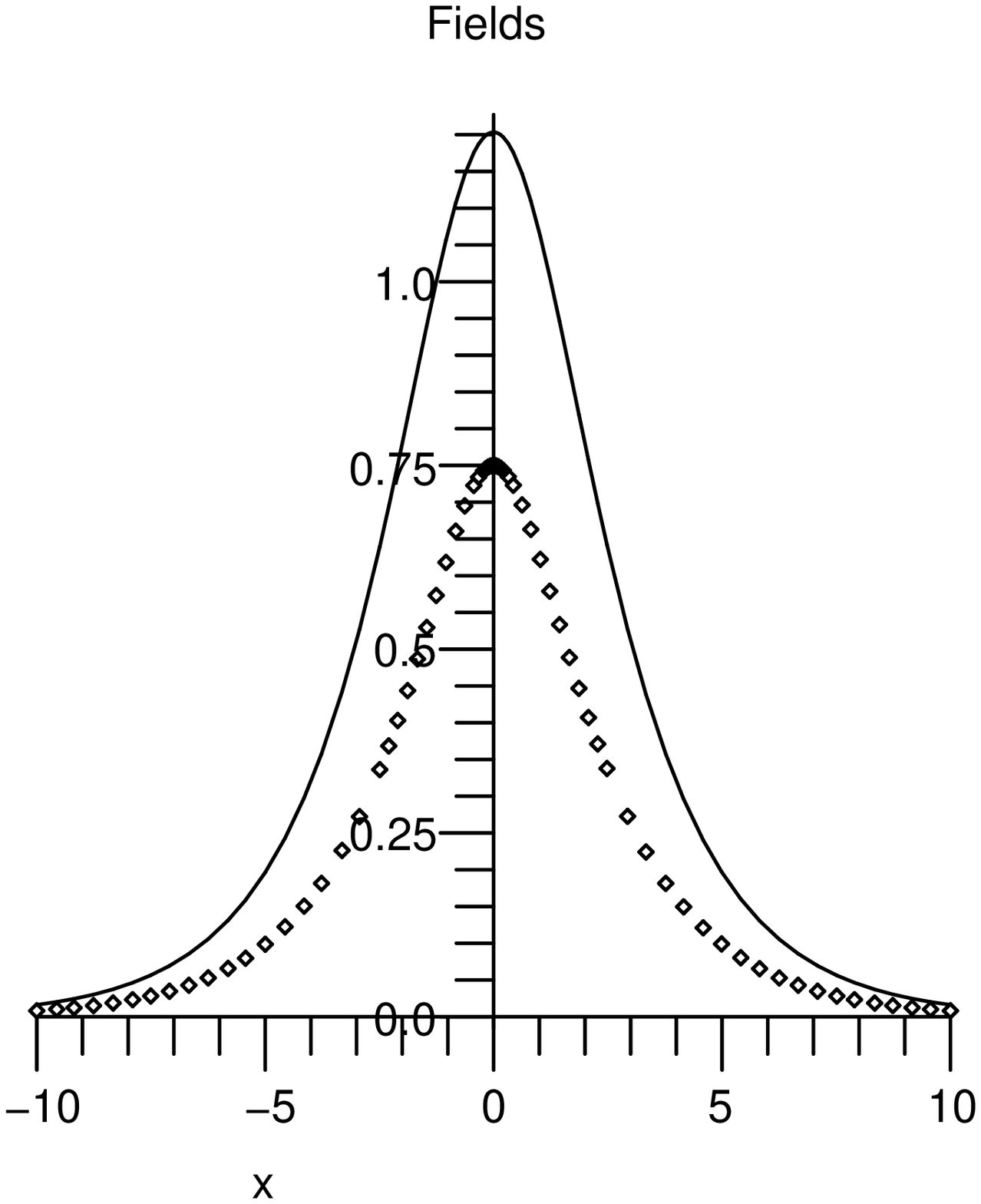,scale=.3}}
\caption{Left hand: Graphs of the fields $\phi_1(x)$ (line) and
$\chi_1(x)$ (point), Middle: Graphs of the fields $\phi_2(x)$ (line)
and $\chi_2(x)$ (point), Right hand: Graphs of the fields
$\phi_3(x)$ (line) and $\chi_3(x)$ (point), all of them are for
$r=0.25$ and $\theta=45$ }
\end{figure}

Fig. (1) shows  that the plots of non-deformed fields $\phi_1$ and
deformed $\chi_1$  which are drawn as  kink. Plots of non-deformed
fields $\phi_2$ and $\phi_3$ and deformed fields $\chi_2$ and
$\chi_3$ are lumps. We see that the graphs of three fields for
non-deformed and deformed are same. In that case they are just
changed by scale.
\begin{figure}[th]
\centerline{\epsfig{file=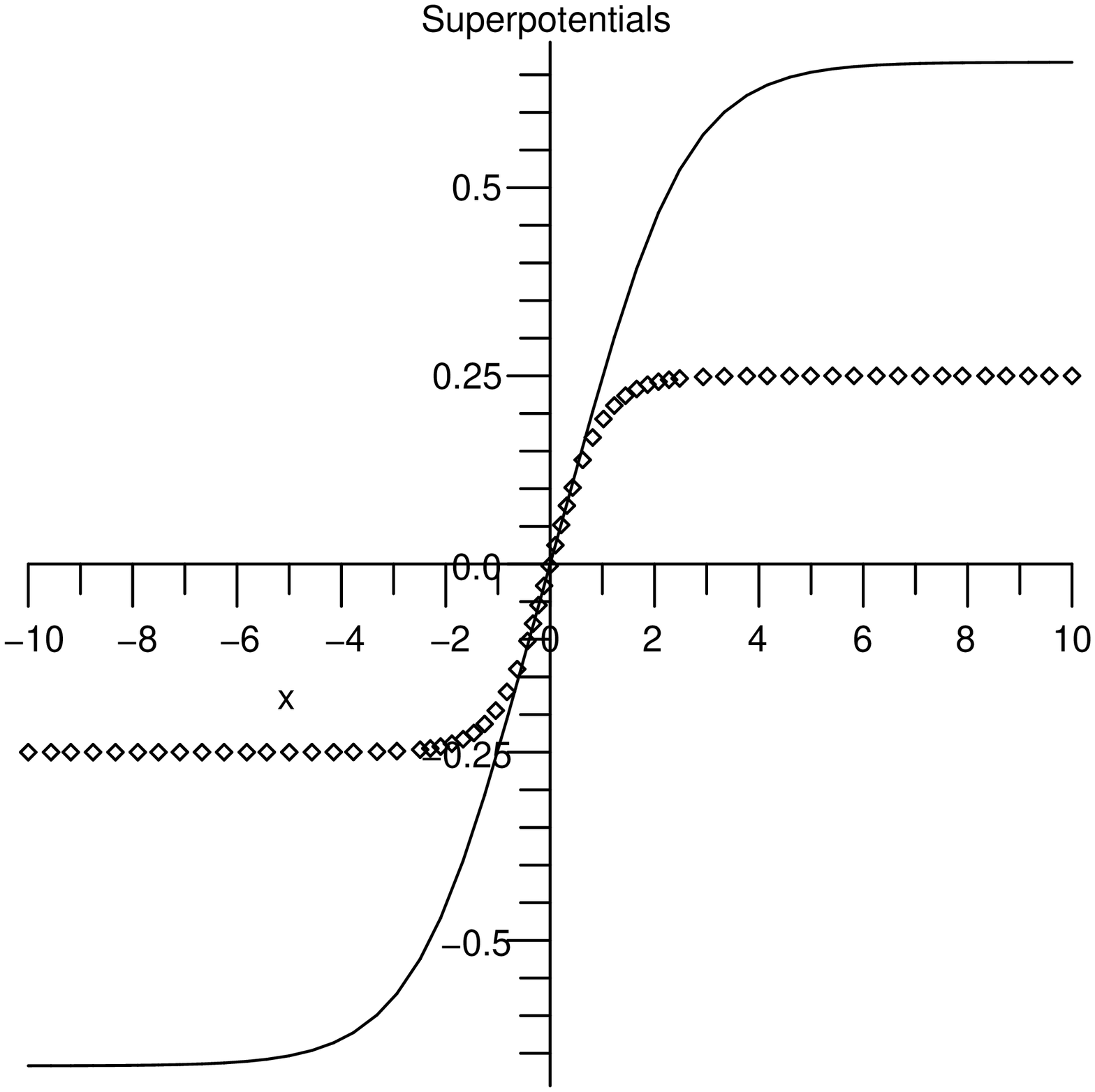,scale=.3}\epsfig{file=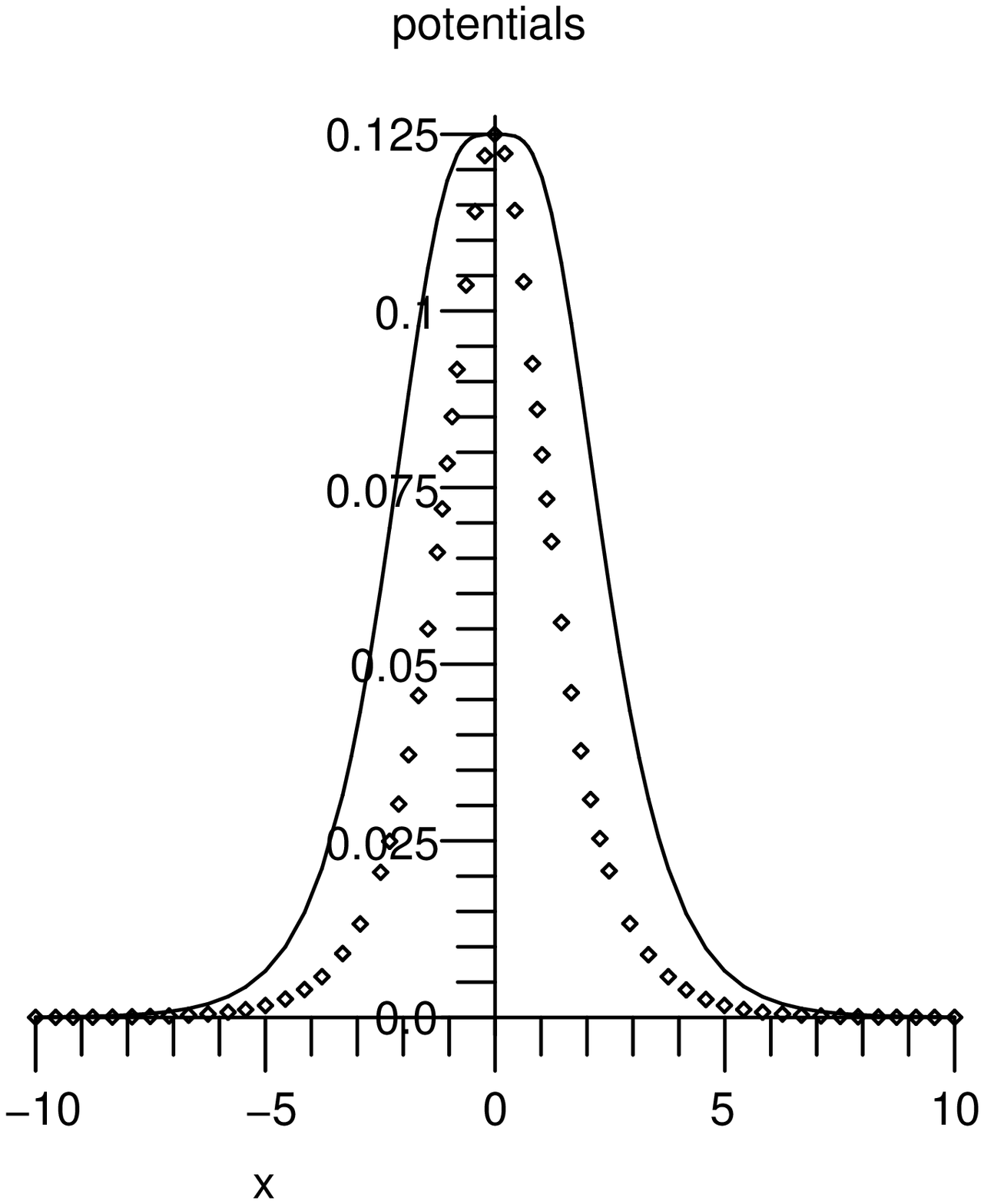,scale=.3}}
\caption{Left hand: Graphs of the fields $\phi_1(x)$ (line) and
$\chi_1(x)$ (point), Middle: Graphs of the fields $\phi_2(x)$ (line)
and $\chi_2(x)$ (point), Right hand: Graphs of the fields
$\phi_3(x)$ (line) and $\chi_3(x)$ (point), all of them are for
$r=0.25$ and $\theta=45$ }
\end{figure}
Fig. (2) shows plots of non-deformed and deformed super-potentials.
Similarly, we see  that the  variation of  two cases are the same,
though by scale are different.

\section{Conclusion}
In this paper, we first have introduced three coupled scalar fields
$\phi_{1}$, $\phi_{2}$ and $\phi_{3}$ [11] and   obtained
topological solutions  by orbit method. Next we have deformed  the
initial fields as $\chi_{1}$, $\chi_{2}$ and $\chi_{3},$ also the
topological solutions for deformed field are obtained by the orbit
method. The solution of deformed and non-deformed three coupled
scalar fields lead us to compare these two cases. So, we have shown
that the variation of fields for two cases with respect to
coordinates in Figs.(1) and (2) are same and different just by
scale. Consequently, different orbit may be lead to different  full
deformed models and solutions. As we know the  deformed fields
similar to non-deformed fields are the form of kink and lump
solutions,  so we have to choose a deformation function such as Eq.
(22). Therefore   we could consider  Eq. (22) in different forms, it
may be interesting for the future work.

\end{document}